\documentclass[showpacs,preprintnumbers,amsmath,amssymb,superscriptaddress,showkeys]{revtex4}
\usepackage{amsmath}
\usepackage{amsfonts}
\usepackage{amssymb}
\usepackage{graphicx}
\usepackage{natbib}
\usepackage[dvipdfm]{hyperref}

\begin{document}
\title{Sudden transitions of trace distance discord of dipole-dipole coupled two qubits}
\author{Zheng-Da Hu}\email{huyuanda1112@jiangnan.edu.cn}
\affiliation{School of Science, Jiangnan University, Wuxi 214122, China}
\author{Jicheng Wang}
\affiliation{School of Science, Jiangnan University, Wuxi 214122, China}
\author{Yixin Zhang}
\affiliation{School of Science, Jiangnan University, Wuxi 214122, China}
\author{Ye-Qi Zhang}
\affiliation{Department of Mathematics and Physics, North China Electric Power University, Beijing 102206, China}
\begin{abstract}
We investigate the exact dynamics of trace distance discord by considering two qubits under dephasing whose states belong to a class of $X$ states beyond Bell diagonal form. The necessary condition for the occurrence of freezing trace distance discord is found and compared with that of entropic discord. For an illustration, we consider two interacting qubits coupled to independent reservoirs and demonstrate these dynamical properties of trace distance discord. It is interesting to find that the freezing trace distance discord exists even for $X$ states without maximally mixed marginals and can be tuned by dipole-dipole coupling of two qubits. Moreover, we consider the initial extended Werner-like states and investigate the differences between trace distance discord and entanglement. The influences of initial state and the dipole-dipole coupling of the two qubits on the dynamics of nonclassical correlations are explored.
\end{abstract}
\maketitle

\section{Introduction}
Entanglement is a kind of nonclassical correlation without classical counterpart and plays a central role in quantum information processing~\cite{nilesen,RevModPhys.81.865}.
However, the entanglement utilized as the quantum resource in the implementation of various quantum tasks is fragile to decoherence, since realistic quantum systems are usually disturbed by their uncontrollable surroundings and then lose coherence.
It has been reported that the entanglement of open quantum systems under decoherence may experience non-asymptotical vanishing although coherence vanishes asymptotically~\cite{PhysRevLett.93.140404}, which is termed as entanglement sudden death (ESD)~\cite{Science.323.598} and has been experimentally demonstrated~\cite{Science.316.579,PhysRevLett.99.180504}.

On the other hand, other type of nonclassical correlation termed as discord~\cite{RevModPhys.84.1655} has been found to beyond entanglement.
The discord can reveal nonclassical behaviors even for unentangled states and be essential in certain quantum computation without entanglement~\cite{PhysRevLett.100.050502,PhysRevLett.101.200501}.
Discord based on entropic quantifiers~\cite{JPA.34.6899,PhysRevLett.88.017901} has been introduced as a measure of nonclassical correlations and attracted much attention (see~\cite{RevModPhys.84.1655} for a comprehensive review).
Nonetheless, it is generally difficult to calculate entropic discord even for two-qubit states and analytical expression can be obtained only for certain special classes of states~\cite{PhysRevA.77.042303,PhysRevA.84.042313,PhysRevA.88.014302}. Alternatively to the entropic approach, quantum correlation can be defined from a unified view based on the idea that the desired correlation is the distance from a given state to the closest state without the desired property~\cite{PhysRevLett.104.080501}.
Then, which distance quantifier is appropriate to define a physically reasonable measure of quantum correlation is of great interest.
For a geometric definition of discord, distance quantifiers based on Schatten $p$-norms have been considered~\cite{PhysRevA.86.024302} and it has been shown that the trace distance based on Schatten $1$-norm (trace norm) is the only suitable distance quantifier among all of these based on Schatten $p$-norms~\cite{PhysRevA.87.064101}.

Compared to other geometric discord based on Schatten $p$-norms including the Schatten $2$-norm case (geometric discord based on Hilbert-Schmidt distance~\cite{PhysRevLett.105.190502}), the trace distance discord (TDD) does not increase under local trace-preserving quantum channels for the unmeasured party~\cite{PhysRevA.86.034101,PhysRevA.87.032340} due to the contractivity of trace distance~\cite{nilesen}.
Besides, for two-qubit systems, the TDD is equivalent to the negativity of quantumness~\cite{PhysRevLett.106.220403,PhysRevA.88.012117,PhysRevLett.110.140501} and exactly computable for an arbitrary $X$ state~\cite{NJP.16.013038}. Furthermore, for two-qubit states of Bell diagonal form, the TDD exhibits remarkable properties under decoherence such as sudden transition~\cite{PhysRevA.87.042115} (this phenomenon is also possessed by other measures of discord~\cite{PhysRevA.88.012120,IntJTheorPhys.53.519,IntJTheorPhys.53.2967}) and double sudden changes~\cite{PhysRevA.87.042115,PhysRevLett.111.250401} (double sudden changes with freezing discord is termed as double sudden transitions in Ref.~\cite{PhysRevLett.111.250401}). Motivated by the available expression of TDD for $X$ states, it is desirable to explore that if the phenomena of sudden transition and double sudden changes with freezing discord exist for $X$ states beyond Bell diagonal form.

In this paper, we study the dynamics of TDD for special $X$ states by considering a dephasing two-qubit system initially prepared to a class of $X$ states beyond Bell diagonal form. The necessary condition for the occurrence of freezing TDD is found and compared with that of entropic discord. Our result shows that the condition of freezing TDD is much weaker than that of freezing entropic discord. For an illustration, we consider two interacting qubits coupled to independent reservoirs and demonstrate these peculiar properties of TDD. It is interesting to find that the phenomena of freezing TDD exist even for the evolving $X$ state without maximally mixed marginals due to the presence of dipole-dipole coupling of the two qubits. By increasing the strength of dipole-dipole coupling, the duration of freezing discord is prolonged in the sudden transition process while it is shortened in the double sudden changes process. Furthermore, we proceed to consider the initial extended Werner-like states and investigate the differences between discord and entanglement. The influences of initial parameters and the dipole-dipole coupling between the two qubits on the dynamics of nonclassical correlations are analyzed. We find that non-Markovian revivals and quantum interference can be induced by the dipole-dipole coupling.

This paper is organized as follows. In Sec.~\ref{sec:sec2}, we introduce the TDD and give the conditions for the occurrence of sudden transition and double sudden changes with freezing discord. In Sec.~\ref{sec:sec3}, we consider an illustrative model as two interacting qubits disturbed by independent reservoirs and demonstrate the phenomena of freezing discord. The effect of dipole-dipole coupling between the two qubits is also discussed. Section~\ref{sec:sec4} is devoted to investigating the differences between TDD and concurrence. The influences of parameters in the EWL state and the dipole-dipole coupling between the two qubits on the dynamics of nonclassical correlations are explored. Conclusions are given at last in Sec.~\ref{sec:sec5}.

\section{Freezing of Trace distance discord}\label{sec:sec2}

In this section, we investigate the dynamics of quantum correlation for a
class of $X$ states with maximally mixed marginals and give the necessary conditions where phenomena of freezing discord occur.
First, we briefly outline basic concepts of trace distance discord (TDD) between a bipartite system (say qubits $A$ and $B$).
Following the idea proposed in Ref.~\cite{PhysRevLett.104.080501}, the TDD of $A$ and $B$ is defined by the trace distance
between $\rho$ and its closest classical-quantum state $\chi_{\rho}$~\cite{PhysRevA.86.024302,PhysRevA.87.064101}, i.e.,
\begin{eqnarray}
D_{G}(\rho)=\min_{\chi_{\rho}\in\Omega_{0}}||\rho-\chi_{\rho}||_{1},
\end{eqnarray}
where $||\Omega||_{1}=\mathrm{Tr}(\sqrt{\Omega^{\dag}\Omega})$ is the Schatten $1$-norm (trace norm)
and $\Omega_{0}$ is the set of classical-quantum states with zero discord. The classical-quantum states $\chi_{\rho}$ can be expressed as
\begin{eqnarray}
\chi_{\rho}=\sum_{k}p_{k}\left\vert \psi_{k}^{A}\left\rangle {}\right\langle
\psi_{k}^{A}\right\vert\otimes\rho_{k}^{B},
\end{eqnarray}
where $\{p_{k}\}$ is a set of statistical probability distribution with $0\leq
p_{k}\leq1$ and $\sum_{k}p_{k}=1$, $\{\left\vert \psi_{k}^{A}\right\rangle \}$
is a complete set of orthonormal basis of subsystem $A$, $\rho_{k}^{B}$ is the $k$th
general quantum state of subsystem $B$ with $k=1,2$ for the qubit case. It is worth noting that other measures of geometric discord based on other Schatten $p$-norms~\cite{PhysRevA.86.024302} including the geometric discord based on Hilbert-Schmidt distance~\cite{PhysRevLett.105.190502}) have also been proposed.  However, it has been shown that the TDD based on Schatten $1$-norm is the only suitable quantum correlation measure~\cite{PhysRevA.87.064101}, which does not increase under local trace-preserving quantum channels for the unmeasured party~\cite{PhysRevA.86.034101,PhysRevA.87.032340} due to the contractivity of trace distance~\cite{nilesen}.

It has been reported in Ref.~\cite{NJP.16.013038} that the TDD is analytically
computable for any two-qubit $X$ state of $X$-shaped matrix form in the
computational basis $\{\left\vert 00\right\rangle ,\left\vert 01\right\rangle
,\left\vert 10\right\rangle ,\left\vert 11\right\rangle \}$ given by
\begin{eqnarray}\label{Xstate}
\rho_{X}=\left(
\begin{array}
[c]{cccc}%
\rho_{11}^{} &  &  & \rho_{14}\\
& \rho_{22} & \rho_{23} & \\
& \rho_{23}^{\ast} & \rho_{33} & \\
\rho_{14}^{\ast} &  &  & \rho_{44}%
\end{array}
\right),
\end{eqnarray}
which subjects to the constraints $\sum_{i=1}^{4}\rho_{ii}^{{}}=1$, $\rho
_{11}\rho_{44}\geq|\rho_{14}|^{2}$ and $\rho_{22}\rho_{33}\geq|\rho_{23}%
|^{2}$. To derive the analytical expression of TDD, one can get rid of phase
factors $\mathrm{e}^{\mathrm{i}\arg\rho_{14}}$ and $\mathrm{e}^{\mathrm{i}%
\arg\rho_{23}}$ through local unitary operations which do not affect the TDD~\cite{NJP.16.013038} to obtain real density matrix
\begin{eqnarray}\label{realx}
\rho_{X}^{'}=\left(
\begin{array}
[c]{cccc}%
\rho_{11}^{} &  &  & |\rho_{14}|\\
& \rho_{22} & |\rho_{23}| & \\
& |\rho_{23}| & \rho_{33} & \\
|\rho_{14}| &  &  & \rho_{44}%
\end{array}
\right).
\end{eqnarray}

\begin{figure}[!htb]
\centering
\includegraphics[angle=0,width=5.5cm]{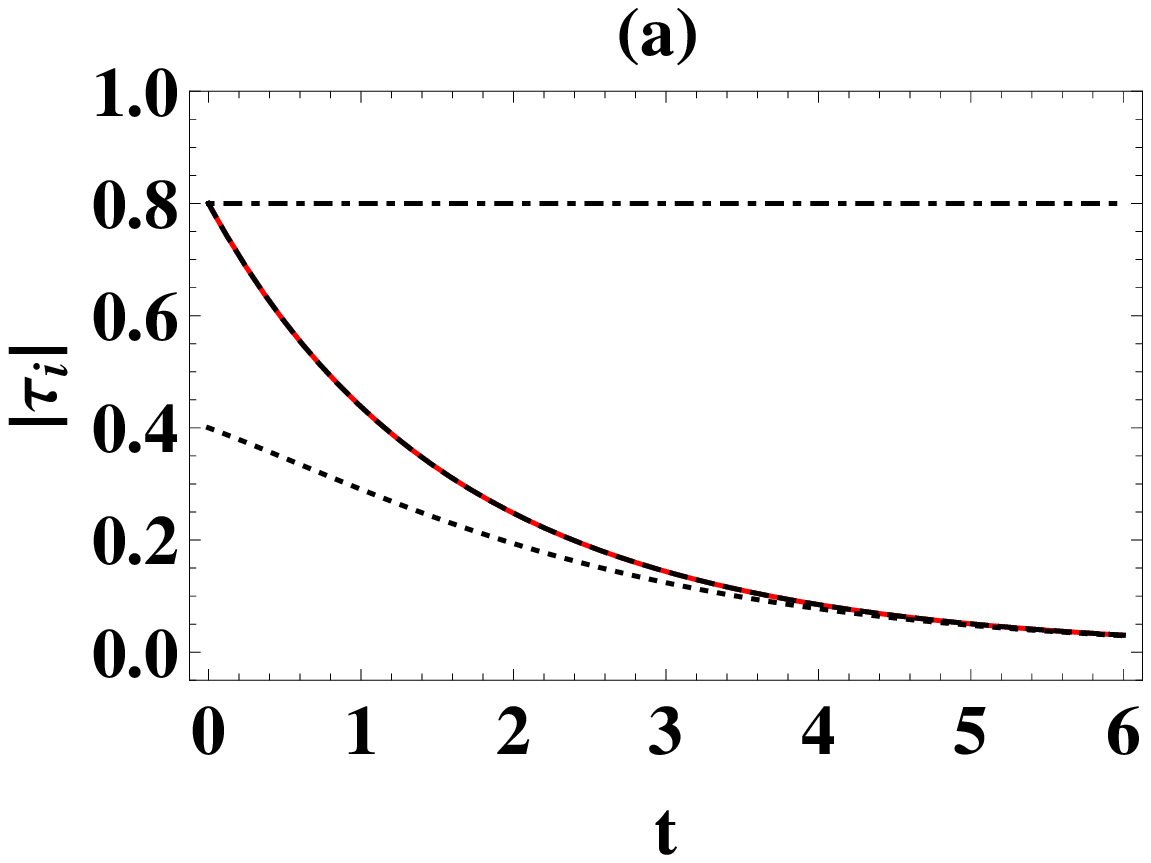}
\includegraphics[angle=0,width=5.5cm]{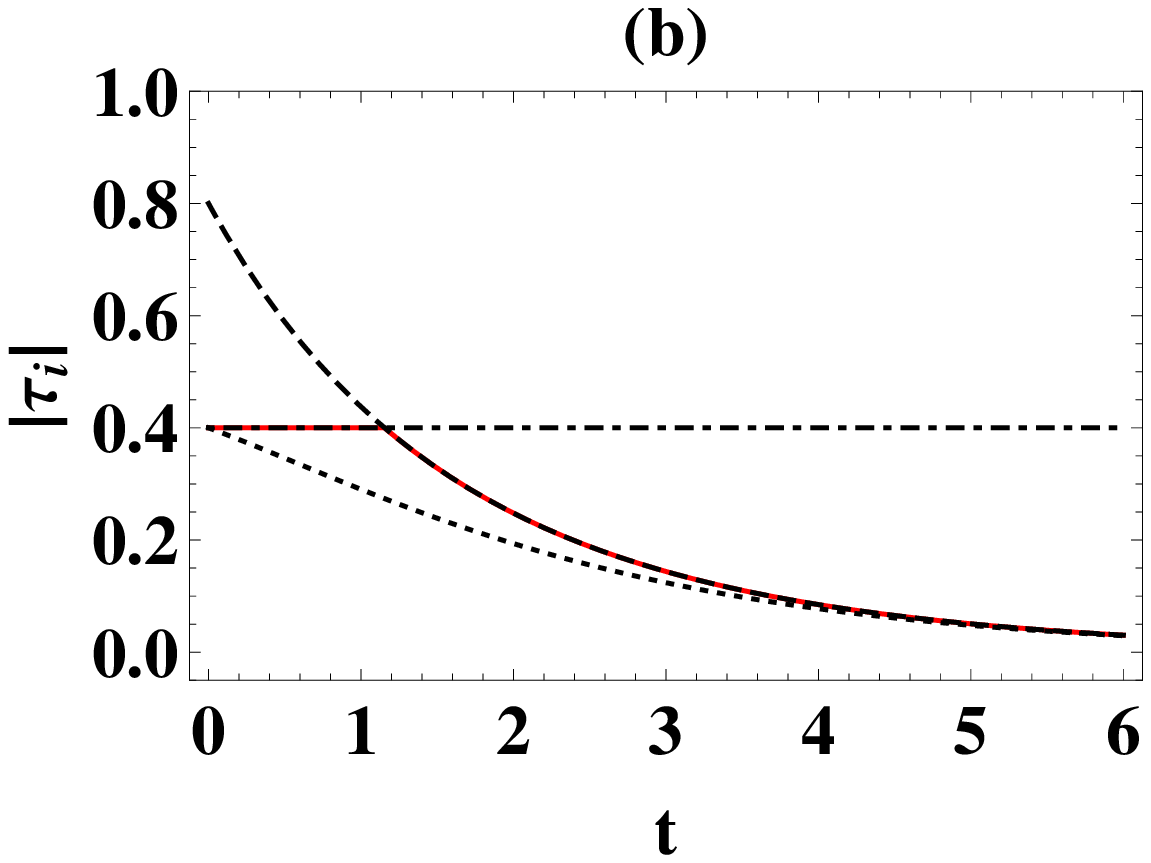}
\includegraphics[angle=0,width=5.5cm]{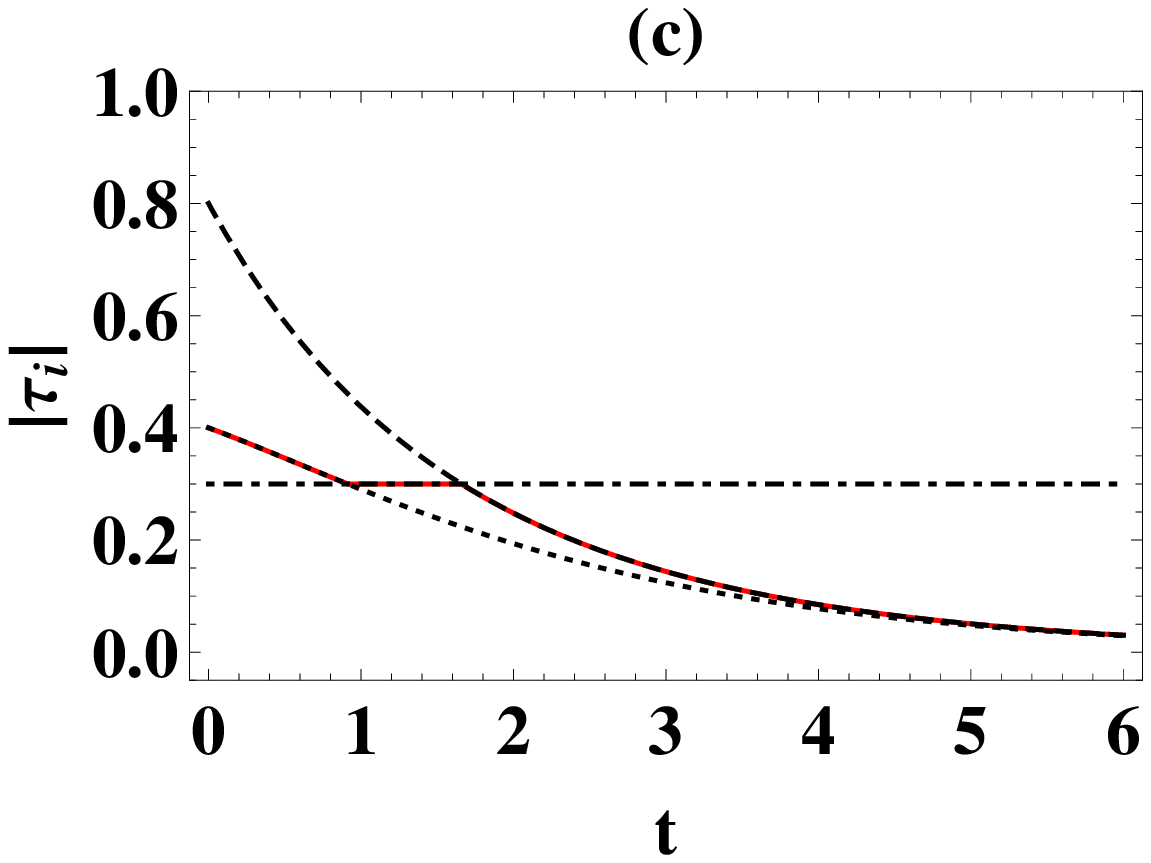}
\caption{(Color online) Dynamics of correlation functions $|\tau_{1}|$ (black-dashed curve), $|\tau_{2}|$ (black-dotted curve), $|\tau_{3}|$ (black-dot-dashed curve) and the TDD (red-solid curve) given by $D_{G}(\rho_{AB}(t))=\mathrm{int}\{|\tau_{1}(t)|,|\tau_{2}(t)|,|\tau_{3}|(t)\}$ for $X$ states (\ref{MMMstate}) with maximally mixed marginals. For convenience, the decoherence factors have been assumed as $d_{1}(t)=\mathrm{e}^{-t}$ and $d_{2}(t)=\mathrm{e}^{-t/2}$ and the initial parameters are specifically set as (a) $c_{1}=0.4$, $c_{2}=c_{3}=0.8$; (b) $c_{1}=c_{3}=0.4$, $c_{2}=0.8$; (c) $c_{1}=0.4$, $c_{2}=0.8$ and $c_{3}=0.3$.}
\label{fig:fig1}
\end{figure}

Next, one can further parameterize state (\ref{realx}) in Bloch representation as
\begin{eqnarray}\label{realxbloch}
\rho_{X}^{'}=&\frac{1}{4}(I\otimes I+\sum_{i=1}^{3}x_{i}\sigma_{i}\otimes I+\sum_{i=1}^{3}y_{i}I\otimes\sigma_{i}+\sum_{i,j=1}^{3}T_{ij}\sigma_{i}\otimes\sigma_{j}),
\end{eqnarray}
where $I$ is the $2\times2$ identity operator, $\sigma_{i}$ $(i=1,2,3)$ are
the standard Pauli matrices, $x_{i}=\mathrm{Tr}[\rho_{X}^{'}(\sigma_{i}\otimes I)]$ and $y_{i}=\mathrm{Tr}[\rho_{X}^{'}(I\otimes\sigma_{i})]$ are the components of the local Bloch vectors $\vec{x}$ and $\vec{y}$ corresponding to the marginal states $\rho_{A}$ and $\rho_{B}$, and $T_{ij}=\mathrm{Tr}[\rho_{X}^{'}(\sigma_{i}\otimes\sigma_{j})]$ are elements of the $3\times3$ real correlation matrix $T$. For state (\ref{realx}), one can easily obtain
\begin{eqnarray}
\vec{x}  & = & (x_{1},x_{2},x_{3})=(0,0,2(\rho_{11}+\rho_{22})-1),\nonumber\\
\vec{y}  & = & (y_{1},y_{2},y_{3})=(0,0,2(\rho_{11}+\rho_{33})-1),\nonumber\\
T  &  = & \left(
\begin{array}
[c]{ccc}%
\tau_{1} &  & \\
& \tau_{2} & \\
&  & \tau_{3}%
\end{array}
\right),
\end{eqnarray}
with $\tau_{1}=2(|\rho_{23}|+|\rho_{14}|)$,
$\tau_{2}=2(|\rho_{23}|-|\rho_{14}|)$ and $\tau_{3}=2(\rho_{11}+\rho_{44})-1$ leading to $|\tau_{1}|\geq|\tau_{2}|$.
Then, the TDD for the two-qubit $X$ state (\ref{Xstate})
can be expressed as~\cite{NJP.16.013038}
\begin{eqnarray}
D_{G}(\rho_{X})=\left\{
\begin{array}
[c]{cc}%
|\tau_{1}|, & |\tau_{2}|\leq|\tau_{1}|\leq|\tau_{3}|\\
\sqrt{\frac{\tau_{1}^{2}\tau_{2}^{'2}-\tau_{2}^{2}\tau_{3}^{2}%
}{\tau_{1}^{2}-\tau_{3}^{2}+x_{3}^{2}}}, & \min\{|\tau_{1}|,\tau_{2}^{'}\}>|\tau_{3}|\\
|\tau_{3}|, & \tau_{2}^{'}\leq|\tau_{3}|<|\tau_{1}|
\end{array}
\right.,
\end{eqnarray}
where $\tau_{2}^{'}=\sqrt{\tau_{2}^{2}+x_{3}^{2}}$.
To explore peculiar phenomena such as sudden transition~\cite{PhysRevA.87.042115,PhysRevLett.111.250401} for the dynamics of
TDD under decoherence, we consider that the dynamics of the two-qubit system
$\rho_{AB}(t)$ is described by a special class of $X$ states with maximally
mixed marginals ($\rho_{A}(t)=\rho_{B}(t)=I/2$). The matrix form of $\rho
_{AB}(t)$ (in the computational basis) is given by Eq.~(\ref{Xstate}) with
\begin{eqnarray}\label{MMMstate}
\rho_{11}(t)&=&\rho_{44}(t)=\frac{1+c_{3}}{4},\nonumber\\
\rho_{22}(t)&=&\rho_{33}(t)=\frac{1-c_{3}}{4},\nonumber\\
\rho_{14}(t)&=&\frac{c_{1}-c_{2}}{4}d_{1}(t),\nonumber\\
\rho_{23}(t)&=&\frac{c_{1}+c_{2}}{4}d_{2}(t),
\end{eqnarray}
where $-1\leq c_{i}\leq1$ $(i=1,2,3)$ are real and time independent parameters
representing the initial condition of the state and $d_{i}(t)$ $(i=1,2)$ are
generally complex and time dependent factors characterizing the decoherence process with
\begin{eqnarray}
|d_{i}(t)|\leq|d_{i}(0)|=1.
\end{eqnarray}
The state (\ref{MMMstate}) is general enough to recover the Bell diagonal states
($d_{1}(t)$ and $d_{2}(t)$ are real and $d_{1}(t)=d_{2}(t)$), the Wener states
($|c_{1}|=|c_{2}|=|c_{3}|=c$ and $d_{1}(t)=d_{2}(t)=1$) and the Bell states (
$|c_{1}|=|c_{2}|=|c_{3}|=d_{1}(t)=d_{2}(t)=1$). It has been reported that the
dynamics of TDD can be frozen during certain time interval under universal Markovian quantum channels such as phase flip, bit
flip, and bit-phase flip channels~\cite{PhysRevA.87.042115,PhysRevA.88.012120}, which have been experimentally
demonstrated~\cite{PhysRevLett.110.140501,PhysRevLett.111.250401} for Bell diagonal states. Here, the $X$ states with maximally
mixed marginals are more general than Bell diagonal states and the decoherence mechanism described by $d_{1}(t)$ and $d_{2}(t)$
is more universal which not only includes all these Markovian channels but also
reveals non-Markovian decoherence channels. It is desirable to find the necessary condition of the occurrence of freezing TDD and compare it with that of the entropic discord~\cite{PhysRevA.80.044102,PhysRevLett.104.200401}.
For the special $X$ state (\ref{MMMstate}) with maximally mixed marginals, one can easily
check that $x_{3}(t)\equiv0$, and express the TDD very simply as follows,
\begin{eqnarray}
D_{G}(\rho_{AB}(t))=\mathrm{int}\{|\tau_{1}(t)|,|\tau_{2}(t)|,|\tau
_{3}(t)|\},
\end{eqnarray}
where%
\begin{eqnarray}\label{3tau}
|\tau_{1}(t)|  &=&|\frac{c_{1}+c_{2}}{2}d_{2}(t)|+|\frac{c_{1}-c_{2}}{2}%
d_{1}(t)|,\nonumber\\
|\tau_{2}(t)|  &=&\vert |\frac{c_{1}+c_{2}}{2}d_{2}(t)|-|\frac
{c_{1}-c_{2}}{2}d_{1}(t)|\vert ,\nonumber\\
|\tau_{3}(t)|  &=&|c_{3}|,
\end{eqnarray}
and "int" denotes the intermediate value. Since $|\tau_{1}(t)|$, $|\tau
_{1}(t)|$ are time dependent with
\begin{eqnarray}
|\tau_{1}(t)|  &  \leq|\tau_{1}(0)|=\max\{|c_{1}|,|c_{2}|\},\nonumber\\
|\tau_{2}(t)|  &  \leq|\tau_{2}(0)|=\min\{|c_{1}|,|c_{2}|\},
\end{eqnarray}
according to Eq.~(\ref{3tau}) and $|\tau_{3}(t)|=|c_{3}|$ is time independent, three
different behaviors of TDD can be concluded as follows: (i) if $|c_{3}%
|\geq\max\{|c_{1}|,|c_{2}|\}$, then $D_{G}(\rho_{AB}(t))=|\tau_{1}(t)|$ which
is always time dependent and therefore no sudden transition exists shown as
Fig.~\ref{fig:fig1}(a); (ii) if $\min\{|c_{1}|,|c_{2}|\}\leq|c_{3}|<\max
\{|c_{1}|,|c_{2}|\}$, then $D_{G}(\rho_{AB}(t))=\min\{|c_{3}|,|\tau_{1}(t)|\}$
where sudden transition occurs shown as Fig.~\ref{fig:fig1}(b); (iii) if
$|c_{3}|<\min\{|c_{1}|,|c_{2}|\}$, then $D_{G}(\rho_{AB}(t))=\mathrm{int}%
\{|\tau_{1}(t)|,|\tau_{2}(t)|,|\tau_{3}(t)|\}$ where double sudden changes
occur which is shown in Fig.~\ref{fig:fig1}(c). As a result, the necessary condition for
the occurrence of sudden transition is $|c_{3}|<\max\{|c_{1}|,|c_{2}|\}$ and
even double sudden changes may occur when the more strict condition
$|c_{3}|<\min\{|c_{1}|,|c_{2}|\}$ is satisfied. It should be pointed out that this condition does not depend on the detail of decoherence, i.e., the concrete forms for $d_{1}(t)$ and $d_{2}(t)$ while the necessary condition for sudden transition of the
traditional entropic discord is that $c_{2}=-c_{1}c_{3}$ and $|d_{1}(t)|=|d_{2}(t)|$~\cite{PLA.376.3011}. As the condition $c_{2}=-c_{1}c_{3}$ and $|d_{1}(t)|=|d_{2}(t)|$ is much more strict than $|c_{3}|<\min\{|c_{1}|,|c_{2}|\}$, the freezing phenomenon is more feasible for the TTD than for the entropic discord. In what follows, we give a concrete physical model to illustrate these dynamical properties of TDD under dephasing.

\section{Dynamics of TDD between two interacting qubits under dephasing for initial $X$ states with maximally mixed marginals}\label{sec:sec3}

\begin{figure}[!htb]
\centering
\includegraphics[angle=0,width=5.5cm]{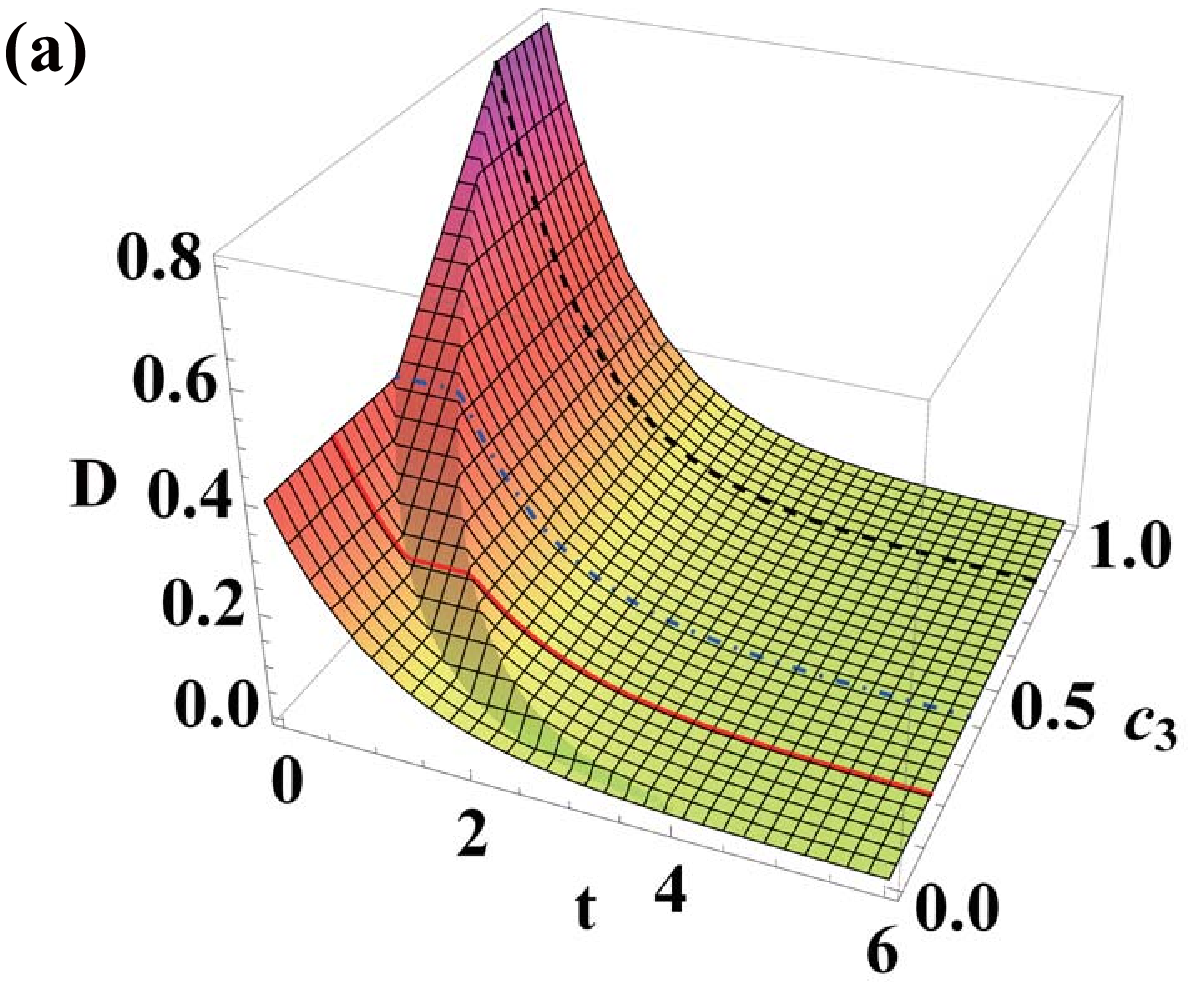}
\includegraphics[angle=0,width=5.5cm]{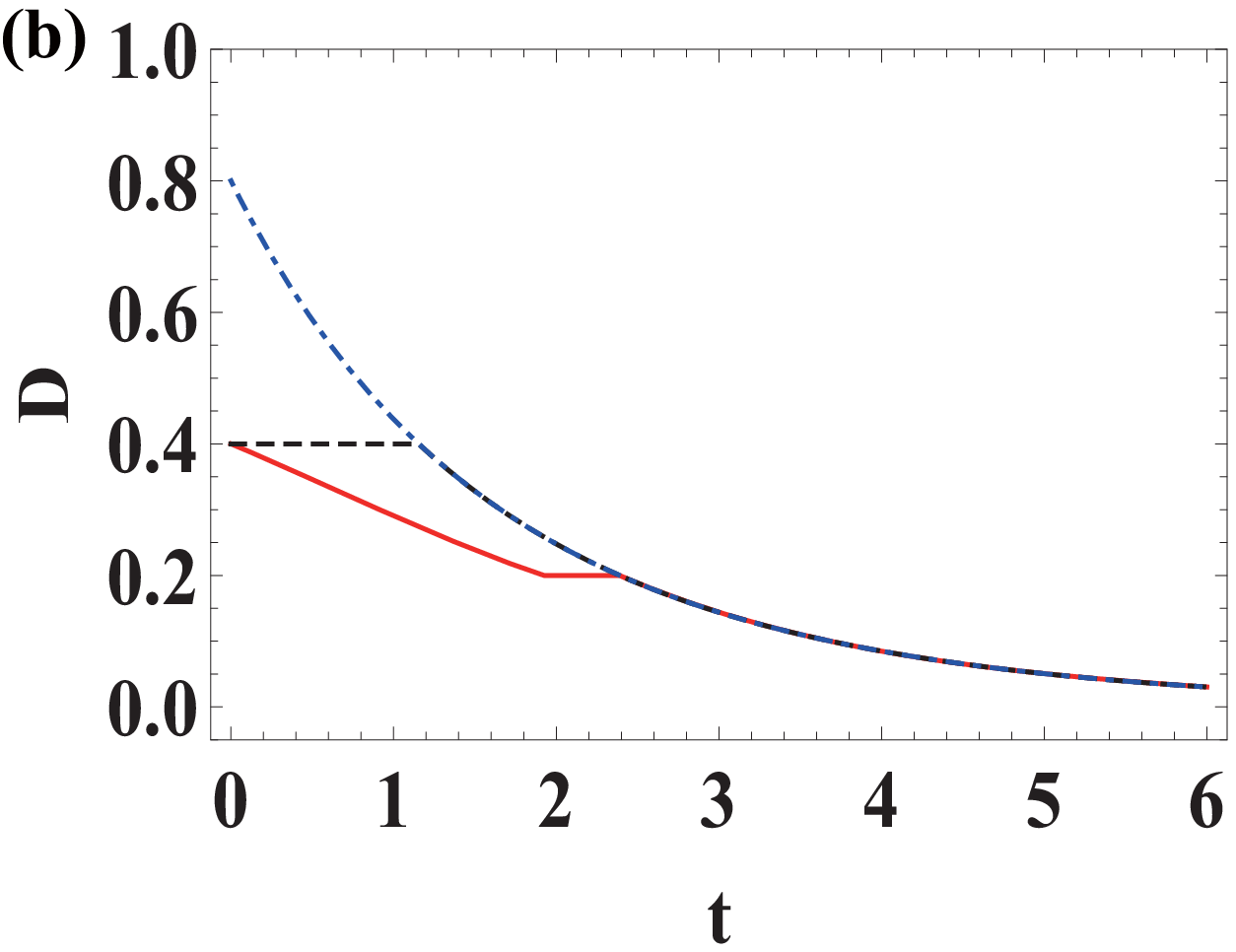}
\caption{(Color online) (a) TDD as a function of $c_{3}$ and $t$ for initial $X$ states with maximally mixed marginals with $c_{1}=0.8$, $c_{2}=0.4$. (b) Dynamics of TDD for specific values of $c_{3}$ with $c_{3}=$ $0.2$ (red-solid curve), $0.4$ (black-dashed curve) and $0.8$ (blue-dot-dashed curve) (also displayed in the $3$D plot of panel (a)). Other parameters are set as $g=0$, $T_{R}=\omega_{0}=1$ and $T_{S}=2$.}
\label{fig:fig2}
\end{figure}

In this section, to illustrate the dynamics of TDD under dephasing,
we consider a quantum system consisting of two interacting qubits, whose
Hamiltonian is given by~\cite{PhysRevA.80.032114}
\begin{eqnarray}
H_{S}=\frac{\omega_{0}}{2}(\sigma_{A}^{z}+\sigma_{B}^{z})+g(\sigma_{A}%
^{+}\sigma_{B}^{-}+\sigma_{A}^{-}\sigma_{B}^{+}),
\end{eqnarray}
where $\omega_{0}$ is the transition frequency of qubit and $g$ describes the
strength of dipole-dipole coupling between the two qubits $A$ and $B$. As realistic systems are usually open to be unavoidably disturbed by their surroundings, we assume that the two-qubit system undergoes a phase relaxation process caused by two
independent reservoirs. The dynamics of the two-qubit system at time $t$ for an initial $X$ state has the form as Eq.~(\ref{Xstate})
with the elements of the density matrix explicitly given by~\cite{PhysRevA.80.032114}
\begin{eqnarray}\label{xsol1}
\rho_{11}(t)  & = & \rho_{11}(0),\,\rho_{44}(t)=\rho_{44}(0),\nonumber\\
\rho_{22}(t)  & = & \mu_{+}+[\mu_{-}\nu_{+}(t)-\Im(\rho_{23}(0))\frac{2g}{d}\sinh(dt)]\mathrm{e}^{-ut},\nonumber\\
\rho_{33}(t)  & = & \mu_{+}-[\mu_{-}\nu_{+}(t)-\Im(\rho_{23}(0))\frac{2g}{d}\sinh(dt)]\mathrm{e}^{-ut},\nonumber\\
\rho_{14}(t)  & = & \rho_{14}(0)\exp(-2\mathrm{i}\omega_{0}t-2t/T_{S}),\nonumber\\
\rho_{23}(t)  & = & \mathrm{i}[\Im(\rho_{23}(0))\nu_{-}(t)+\mu_{-}\frac{2(g+v)}{d}\sinh(dt)]\mathrm{e}^{-ut}\nonumber\\
&\quad&+\Re(\rho_{23}(0))\mathrm{e}^{-2ut},
\end{eqnarray}
where $\Re(\Omega)$ ($\Im(\Omega)$) denotes the real (imaginary) part of $\Omega$ and
\begin{eqnarray}\label{xsol2}
\mu_{\pm} & = & \frac{1}{2}[\rho_{22}(0)\pm\rho_{33}(0)],\nonumber\\
\nu_{\pm}(t) & = & \cosh(dt)\pm\frac{u}{d}\sinh(dt),\nonumber\\
d  & = & \sqrt{u^{2}-4g(g+v)},\nonumber\\
u  & = & 1/[T_{S}(1+4g^{2}T_{R}^{2})],\nonumber\\
v  & = & 2gT_{R}/[T_{S}(1+4g^{2}T_{R}^{2})],
\end{eqnarray}
with $T_{S}$ and $T_{R}$ corresponding to the characteristic relaxation times for the qubit system and the reservoir, respectively.

\begin{figure}[!htb]
\centering
\includegraphics[angle=0,width=5.5cm]{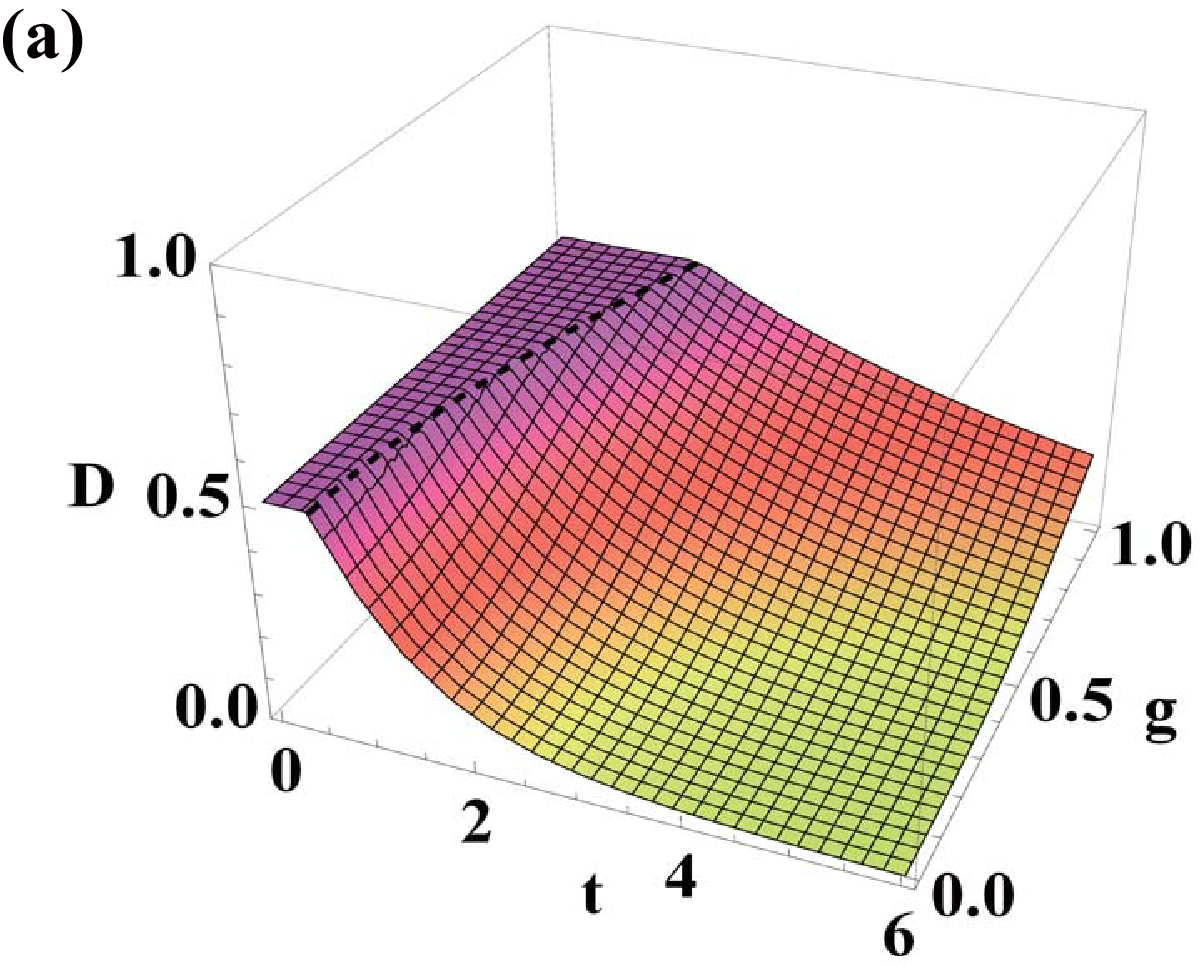}
\includegraphics[angle=0,width=5.5cm]{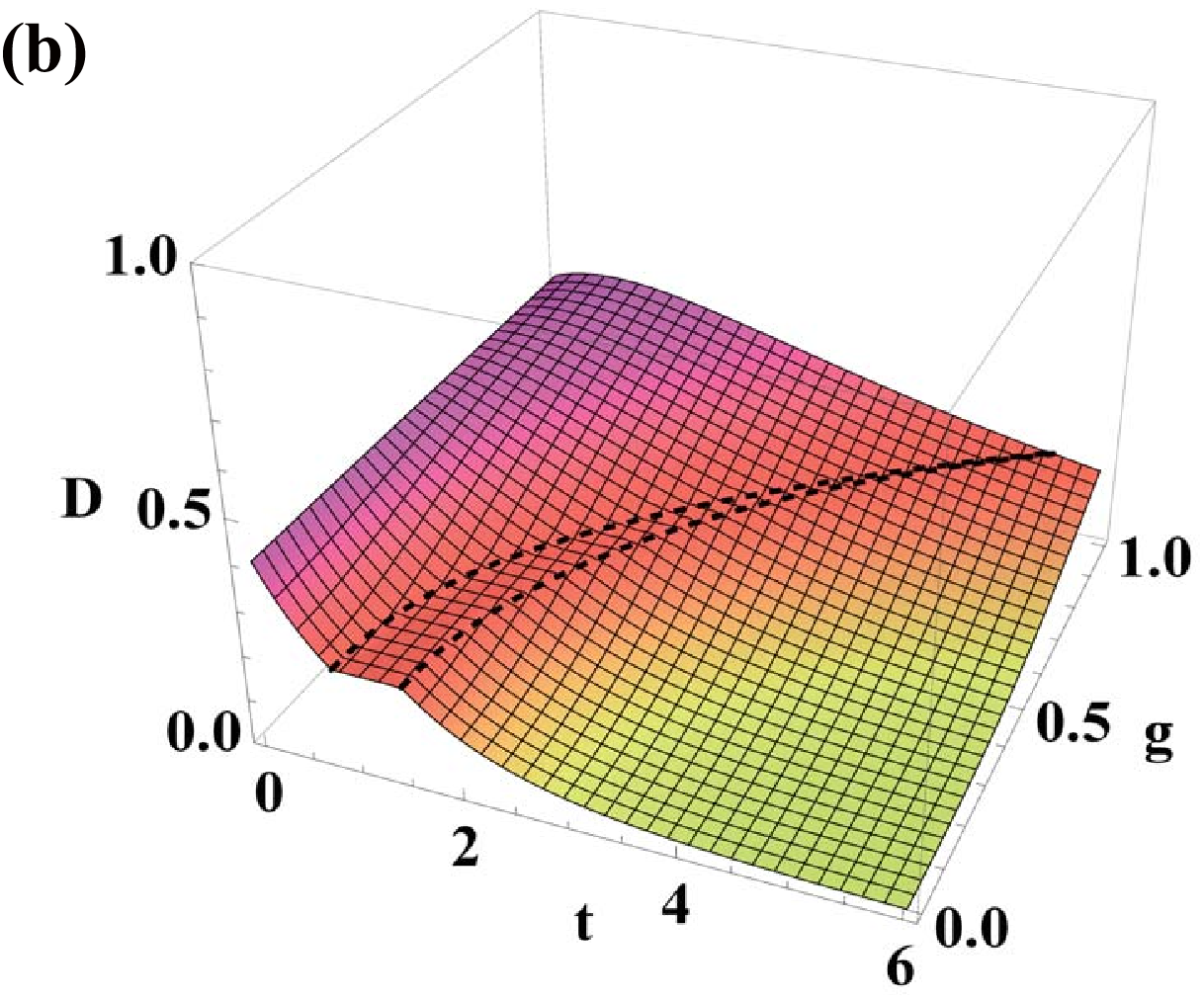}
\caption{(Color online) TDD as a function of $g$ and $t$ for initial $X$ states with maximally mixed marginals with (a) $c_{1}=0.8$ and $c_{2}=c_{3}=0.5$ (b) $c_{1}=0.8$, $c_{2}=0.4$ and $c_{3}=0.2$. The dashed curves correspond to the sudden changes in the decay rate of TDD. Other parameters are set the same as Fig.~\ref{fig:fig2}.}
\label{fig:fig3}
\end{figure}

First, we consider the noninteracting two-qubit case ($g=0$) and the initial state of the two qubits are prepared in a special class of $X$ states with maximally mixed marginals, in terms of which the dynamical evolution of the two-qubit state is still an $X$ state with maximally mixed marginals given by Eq.~(\ref{MMMstate}) with decoherence factors
\begin{eqnarray}
d_{1}(t)&=&\exp(-2\mathrm{i}\omega_{0}t-2t/T_{S}),\nonumber\\
d_{2}(t)&=&\mathrm{e}^{-2ut}.
\end{eqnarray}
According to the discussions in the previous section, the dynamical behavior of TDD depends largely on the relations among the initial parameters $c_{1}$, $c_{2}$ and $c_{3}$. To show this, we fix the values of $c_{1}$ and $c_{2}$ and plot the TDD as a function of $c_{3}$ and $t$ in Fig.~\ref{fig:fig2}(a). It is clearly seen that the TDD exhibits smooth decay, sudden transition and double sudden changes with freezing discord as $c_{3}$ decreases from $1$ to $0$. For the first case $|c_{3}|\geq\max\{|c_{1}|,|c_{2}|\}$, the TDD is an analytic function as
\begin{eqnarray}
D_{G}^{(1)}(\rho_{AB}(t))=|\frac{c_{1}+c_{2}}{2}|\mathrm{e}^{-2ut}+|\frac{c_{1}-c_{2}}{2}|\mathrm{e}^{-\frac{2t}{T_{S}}},
\end{eqnarray}
which monotonously decays with time $t$. For the second case $\min\{|c_{1}|,|c_{2}|\}\leq|c_{3}|<\max\{|c_{1}|,|c_{2}|\}$, the TDD is given by
\begin{eqnarray}
D_{G}^{(2)}(\rho_{AB}(t))=\min\{|c_{3}|,D_{G}^{(1)}(\rho_{AB}(t))\},
\end{eqnarray}
in terms of which sudden transition is indicated and the transition point $t_{\mathrm{c}}$ satisfies to the condition $|c_{3}|=D_{G}^{(1)}(\rho_{AB}(t_{\mathrm{c}}))$. We derive $t_{\mathrm{c}}$ explicitly as
\begin{eqnarray}
t_{\mathrm{c}}=\frac{T_{S}}{2}\ln\frac{\max\{|c_{1}|,|c_{2}|\}}{|c_{3}|},
\end{eqnarray}
which is proportional to the relaxation time $T_{S}$ of qubit. For the third case $|c_{3}|<\min\{|c_{1}|,|c_{2}|\}$, the TDD reads as
\begin{eqnarray}
D_{G}^{(3)}(\rho_{AB}(t))=\mathrm{int}\{|c_{3}|,D_{G}^{(0)}(\rho_{AB}(t)),D_{G}^{(1)}(\rho_{AB}(t))\},
\end{eqnarray}
where we have denoted $\vert|\frac{c_{1}+c_{2}}{2}|\mathrm{e}^{-2ut}-|\frac{c_{1}-c_{2}}{2}|\mathrm{e}^{-\frac{2t}{T_{S}}}\vert=D_{G}^{(0)}(\rho_{AB}(t))$ for convenience. Then, double sudden changes of TDD arise with the first and second change points $t_{c}^{(1)}$ and $t_{c}^{(2)}$ subject to $|c_{3}|=D_{G}^{(0)}(\rho_{AB}(t))$ and $|c_{3}|=D_{G}^{(1)}(\rho_{AB}(t))$, respectively. We can also obtain $t_{c}^{(1)}$ and $t_{c}^{(2)}$ explicitly as
\begin{eqnarray}
t_{\mathrm{c}}^{(1)}&=&\frac{T_{S}}{2}\ln\frac{\min\{|c_{1}|,|c_{2}|\}}{|c_{3}|},\nonumber\\
t_{\mathrm{c}}^{(2)}&=&t_{\mathrm{c}}.
\end{eqnarray}
It is worth noting that the second change point $t_{\mathrm{c}}^{(2)}$ is actually the transition point, near which the transition from classical decoherence to quantum decoherence occurs while the first change point $t_{\mathrm{c}}^{(1)}$ merely means a sudden change of the decay rate of TDD, which has already been indicated in Ref.~\cite{PhysRevLett.111.250401}. The effect of the dipole-dipole coupling $g$ between qubits $A$ and $B$ is also displayed in Fig.~\ref{fig:fig3}. It is interesting to find that the phenomena of sudden transition and double sudden changes with freezing discord still exist even when the evolutional state does not preserve maximally mixed marginals, which can be seen from Eqs.~(\ref{xsol1}) and (\ref{xsol2}) for $g\neq 0$. Moreover, the duration of freezing TDD is prolonged for the sudden transition case when $g$ increases (shown in Fig.~\ref{fig:fig3}(a)) while it is shortened for the double sudden changes case (shown in Fig.~\ref{fig:fig3}(b)).

\begin{figure}[!htb]
\centering
\includegraphics[angle=0,width=5.5cm]{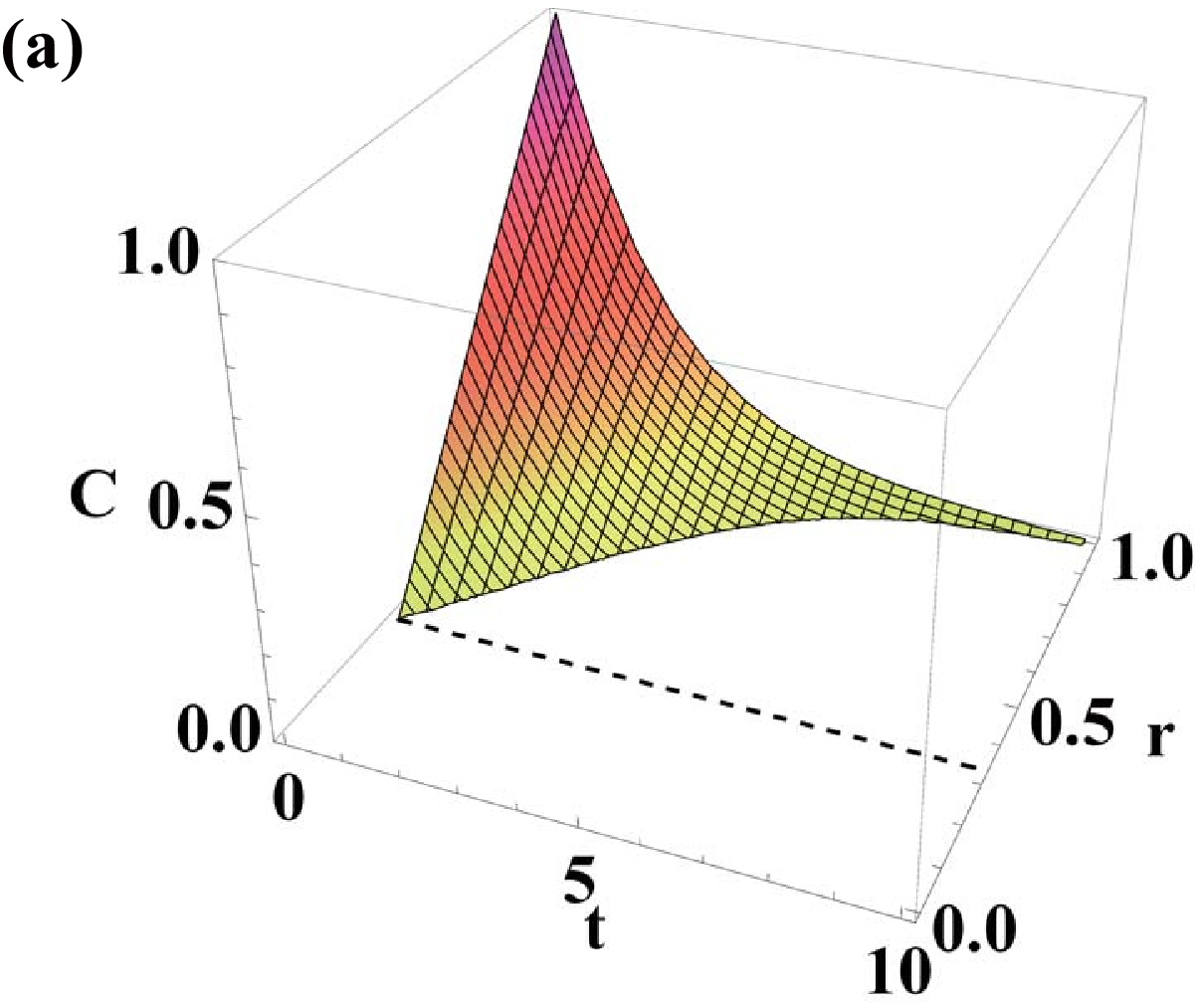}
\includegraphics[angle=0,width=5.5cm]{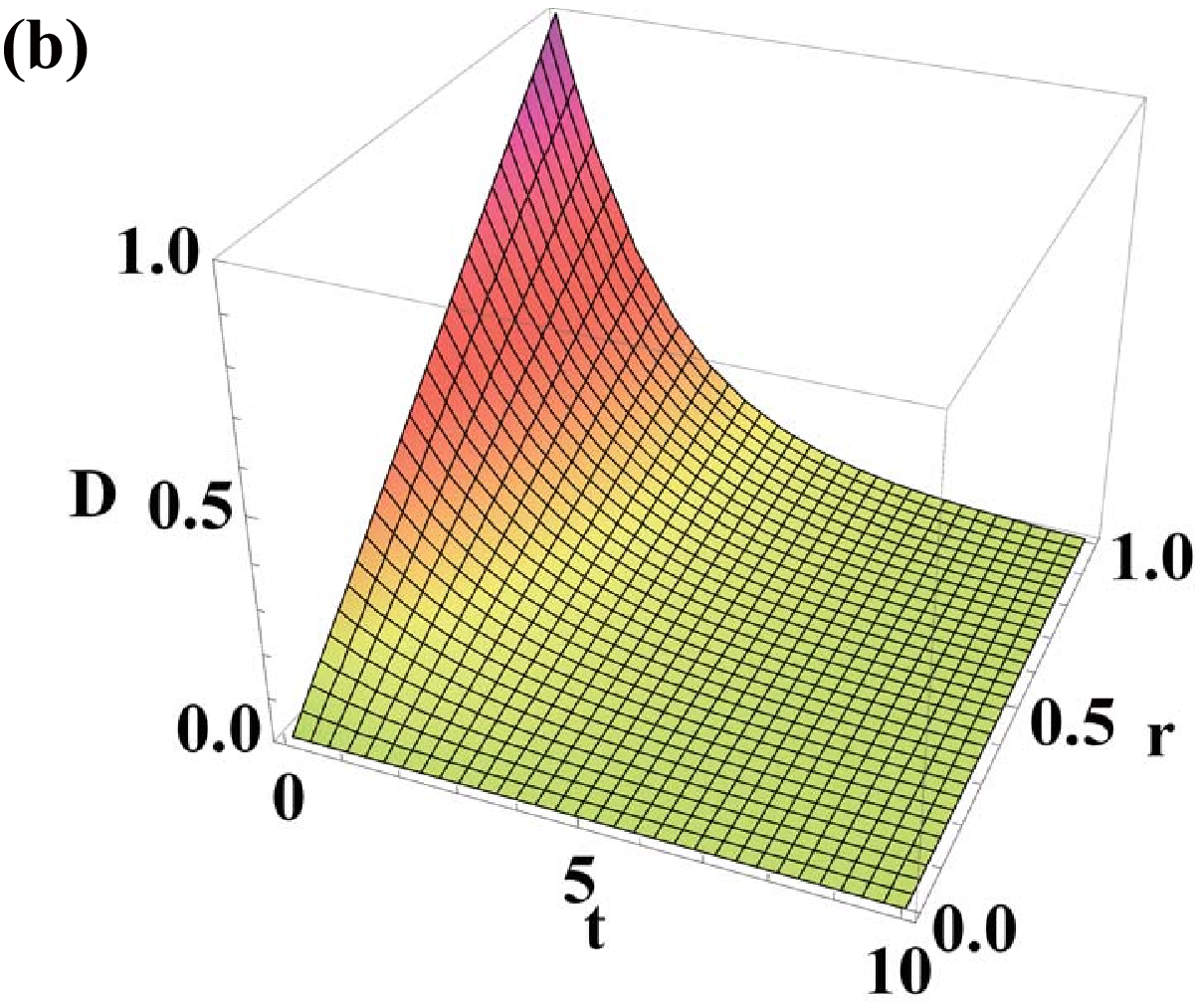}
\caption{(Color online) (a) Concurrence and (b) TDD as functions of $r$ and $t$ for initial EWL states (\ref{EWLPsi0}) with $\theta=\pi/2$ and $\phi=0$. Other parameters are set as $g=1/2$, $T_{R}=\omega_{0}=1$ and $T_{S}=2$.}
\label{fig:fig4}
\end{figure}

\section{Dynamics of TDD between two interacting qubits under dephasing for initial extended Werner-like states}\label{sec:sec4}

In this section, we focus on another special class of $X$ states, i.e., the extended Werner-like states (EWLs) and explore the differences between TDD and entanglement. The measure of entanglement we used is Wootters' concurrence $C$~\cite{PhysRevLett.80.2245}, which varies from $C=0$ for a separable state to $C=1$ for a maximally entangled state. For a two-qubit state $\rho_{AB}$, the concurrence is given by
\begin{equation}
C(\rho_{AB})=\max\{\sqrt{\lambda_{1}}-\sqrt{\lambda_{2}}-\sqrt{\lambda_{3}}-\sqrt{\lambda_{4}},0\},
\end{equation}
where $\lambda_{i}$ ($i=1,2,3,4$) are the eigenvalues, in decreasing order, of the matrix
\begin{equation}
R=\rho_{AB}(\sigma_{y}\otimes\sigma_{y})\rho_{AB}^{\ast}(\sigma_{y}\otimes\sigma_{y}),
\end{equation}
with the asterisk denoting the complex conjugation. Moreover, the concurrence for the two-qubit $X$ state (\ref{Xstate}) can be explicitly expressed as
\begin{equation}\label{xcon}
C(\rho_{AB})=2\max\{0,\vert\rho_{14}\vert-\sqrt{\rho_{22}\rho_{33}},\,
\vert\rho_{23}\vert-\sqrt{\rho_{11}\rho_{44}}\}.
\end{equation}

Then, we consider the qubits $A$ and $B$ are initially prepared in an EWL state, defined by
\begin{equation}\label{EWL}
\rho_{\mathrm{EWL}}^{}=\frac{1-r}{4}I+r\left\vert \xi\right\rangle
\left\langle \xi\right\vert,
\end{equation}
where $0\leq r\leq1$ is the purity of the state, $I$ is the $4\times 4$ identity operator, and $\left\vert \xi\right\rangle
=\left\vert \Psi\right\rangle $ or $\left\vert \Phi\right\rangle $ is the Bell-like state with $\left\vert \Psi\right\rangle =\cos\frac{\theta}{2}\left\vert01\right\rangle +\sin\frac{\theta}{2}\mathrm{e}^{\mathrm{i}\phi}\left\vert 10\right\rangle $, $\left\vert
\Phi\right\rangle =\cos\frac{\theta}{2}\left\vert00\right\rangle +\sin\frac{\theta}{2}\mathrm{e}^{\mathrm{i}\phi}\left\vert 11\right\rangle$, $0\leq\theta<\pi$ and $0\leq\phi<2\pi$. The EWL state reduces to the Werner state when the Bell-like state $\left\vert \xi\right\rangle $ reduces to the maximally entangled Bell state at $\theta=\pi/2$. An important property of EWL state $\rho_{\mathrm{EWL}}^{}$ is that it is entangled for $1/(2\sin\theta+1)<r\leq1$ and disentangled for $0\leq r\leq1/(2\sin\theta+1)$, indicating that the Werner state is entangled only for $1/3<r\leq1$. Since the dipole-dipole coupling Hamiltonian $H_{\mathrm{d-d}}=g(\sigma_{A}^{+}\sigma_{B}^{-}+\sigma_{A}^{-}\sigma_{B}^{+})$ has no effect on $\Phi$-type EWL states $\rho^{(\Phi)}$, i.e., $[H_{\mathrm{d-d}},\rho^{(\Phi)}]=0$, we would like to pay attention to initial $\Psi$-type EWL states $\rho^{(\Psi)}(0)$ with the elements of density matrix
\begin{eqnarray}
\rho_{11}^{(\Psi)}(0)&=&\rho_{44}^{(\Psi)}(0)=\frac{1-r}{4},\nonumber\\
\rho_{22}^{(\Psi)}(0)&=&\frac{1-r}{4}+r\cos^{2}\frac{\theta}{2},\nonumber\\
\rho_{33}^{(\Psi)}(0)&=&\frac{1-r}{4}+r\sin^{2}\frac{\theta}{2},\nonumber\\
\rho_{14}^{(\Psi)}(0)&=&0,\,\rho_{23}^{(\Psi)}(0)=\frac{r}{2}\sin\theta\mathrm{e}^{-\mathrm{i}\phi},
\end{eqnarray}
in terms of which the concurrence and TDD at the initial time are given by
\begin{eqnarray}\label{EWLPsi0}
C(\rho^{(\Psi)}(0))&=&\max\{0,r|\sin\theta|-\frac{1-r}{2}\},\nonumber\\
D_{G}(\rho^{(\Psi)}(0))&=&r|\sin\theta|.
\end{eqnarray}

\begin{figure}[!htb]
\centering
\includegraphics[angle=0,width=5.5cm]{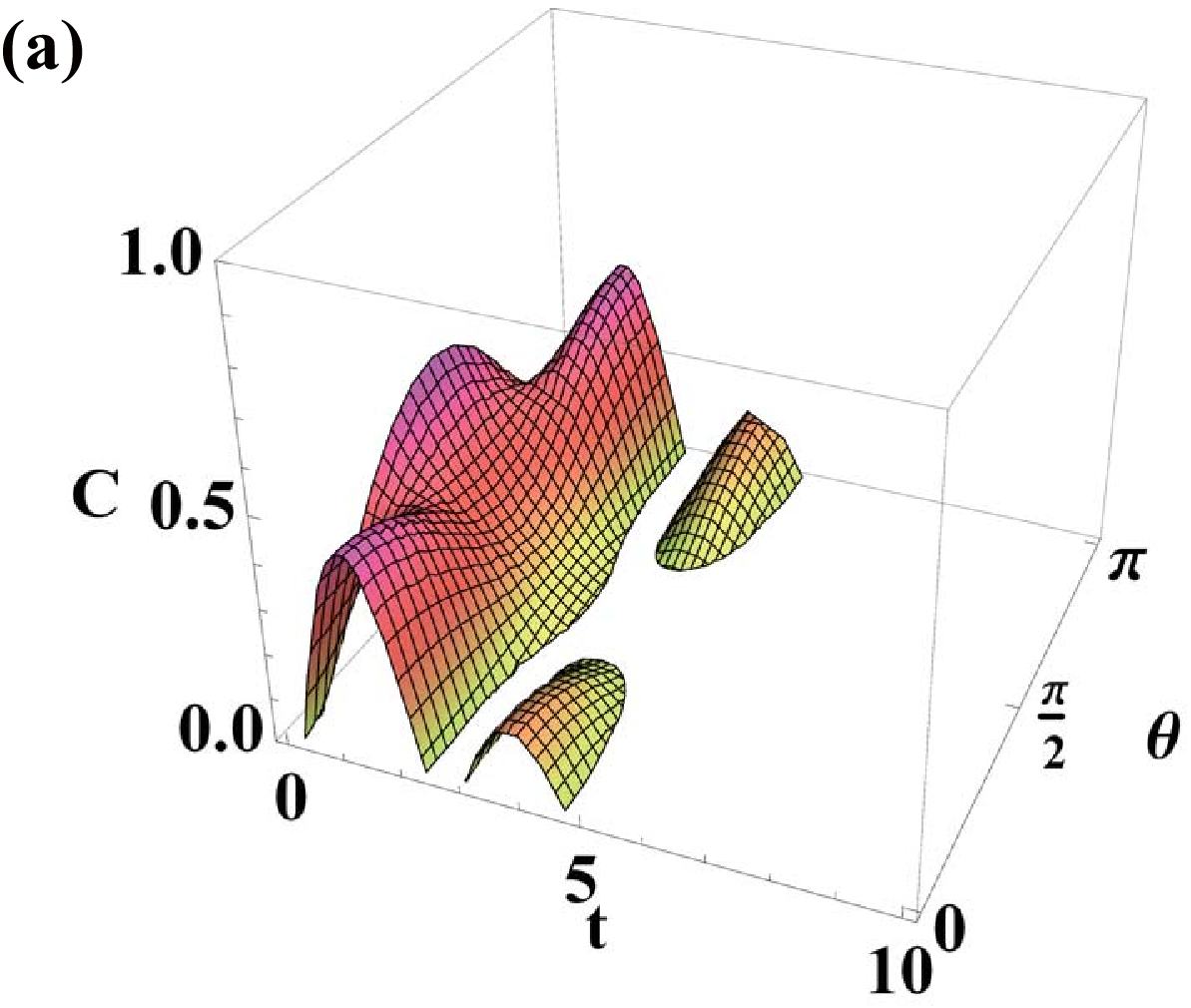}
\includegraphics[angle=0,width=5.5cm]{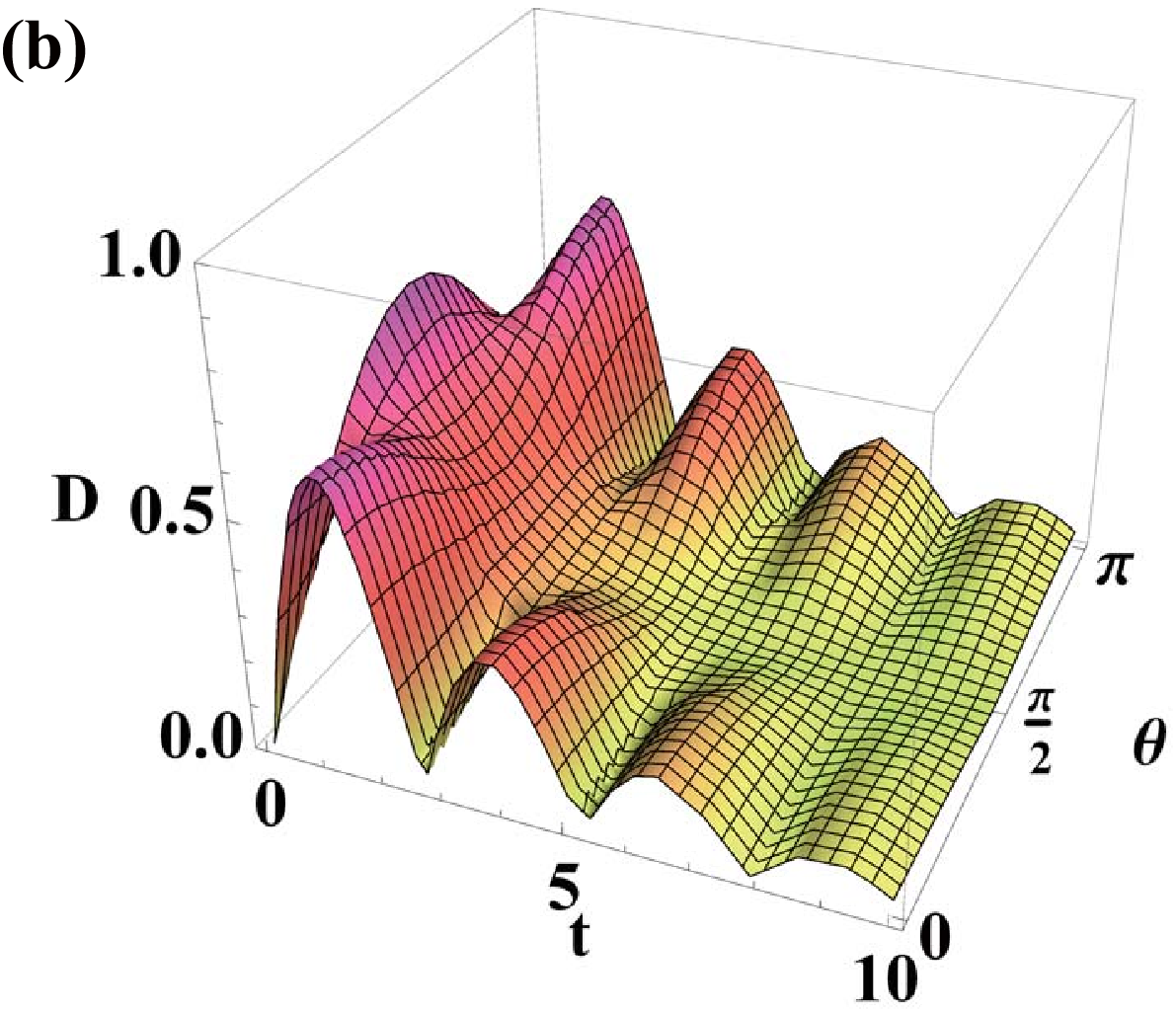}
\caption{(Color online) (a) Concurrence and (b) TDD as functions of $\theta$ and $t$ for initial EWL states (\ref{EWLPsi0}) with $r=2/3$ and $\phi=0$. Other parameters are set the same as Fig.~\ref{fig:fig4}.}
\label{fig:fig5}
\end{figure}

First, we explore the influence of the purity $r$ of the EWL state on the nonclassical correlations and display the concurrence and TDD as functions of $r$ and $t$ in Fig.~\ref{fig:fig4}. Two qualitative differences can be observed by comparing Fig.~\ref{fig:fig4}(a) with  Fig.~\ref{fig:fig4}(b). For initial parameters $\theta=\pi/2$ and $\phi=0$, it is clearly shown that the initial concurrence is zero for $r\leq 1/3$ while the TDD is always nonzero, which means discord contains certain nonclassicality that is missed by entanglement. In this sense, the nonclassical correlation quantified by discord is more universal than that quantified by entanglement. Another significant difference is that the concurrence decays with sudden vanishing at finite time (ESD)~\cite{Science.323.598} while the TDD decays asymptotically to zero.

\begin{figure}[!htb]
\centering
\includegraphics[angle=0,width=5.5cm]{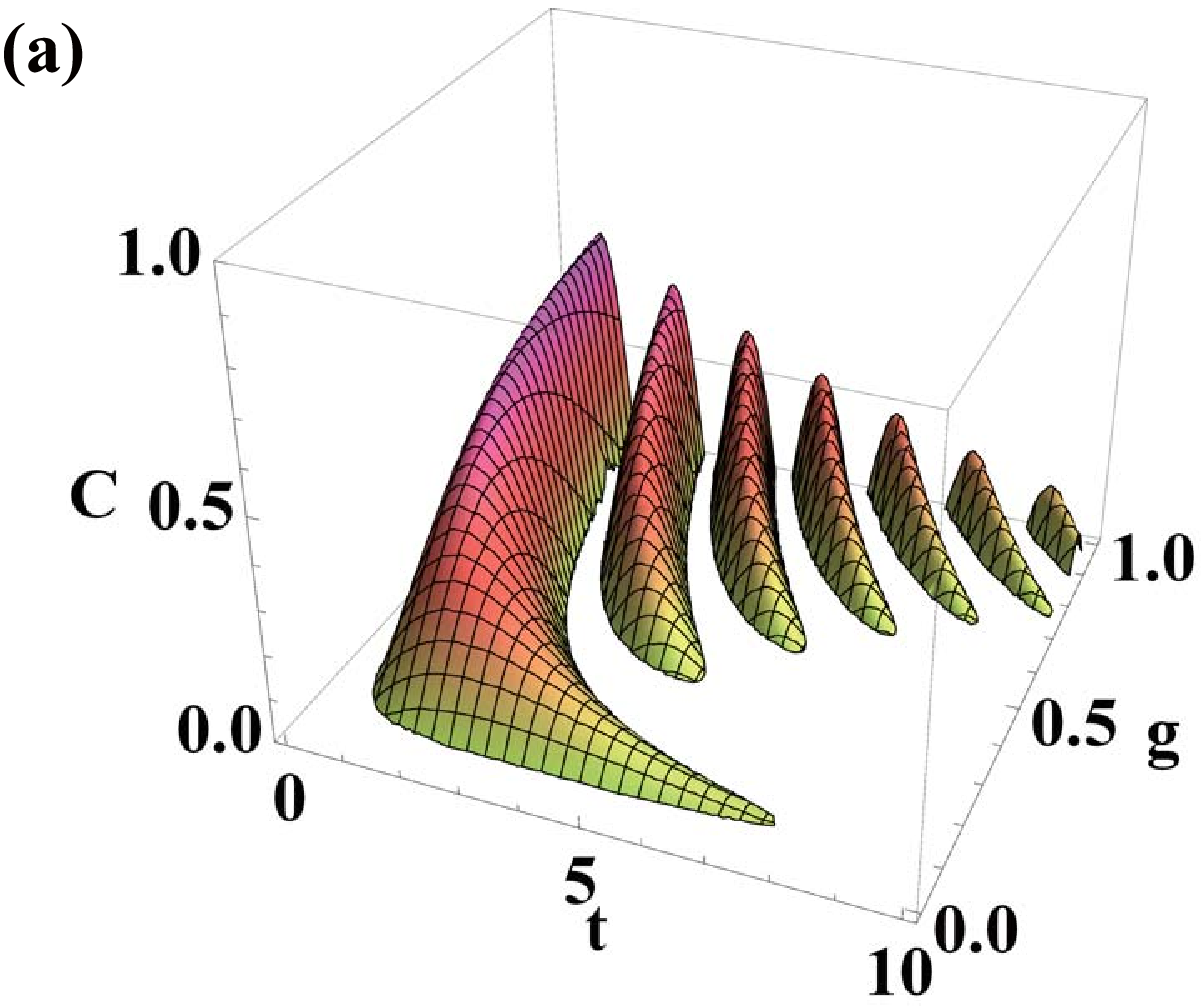}
\includegraphics[angle=0,width=5.5cm]{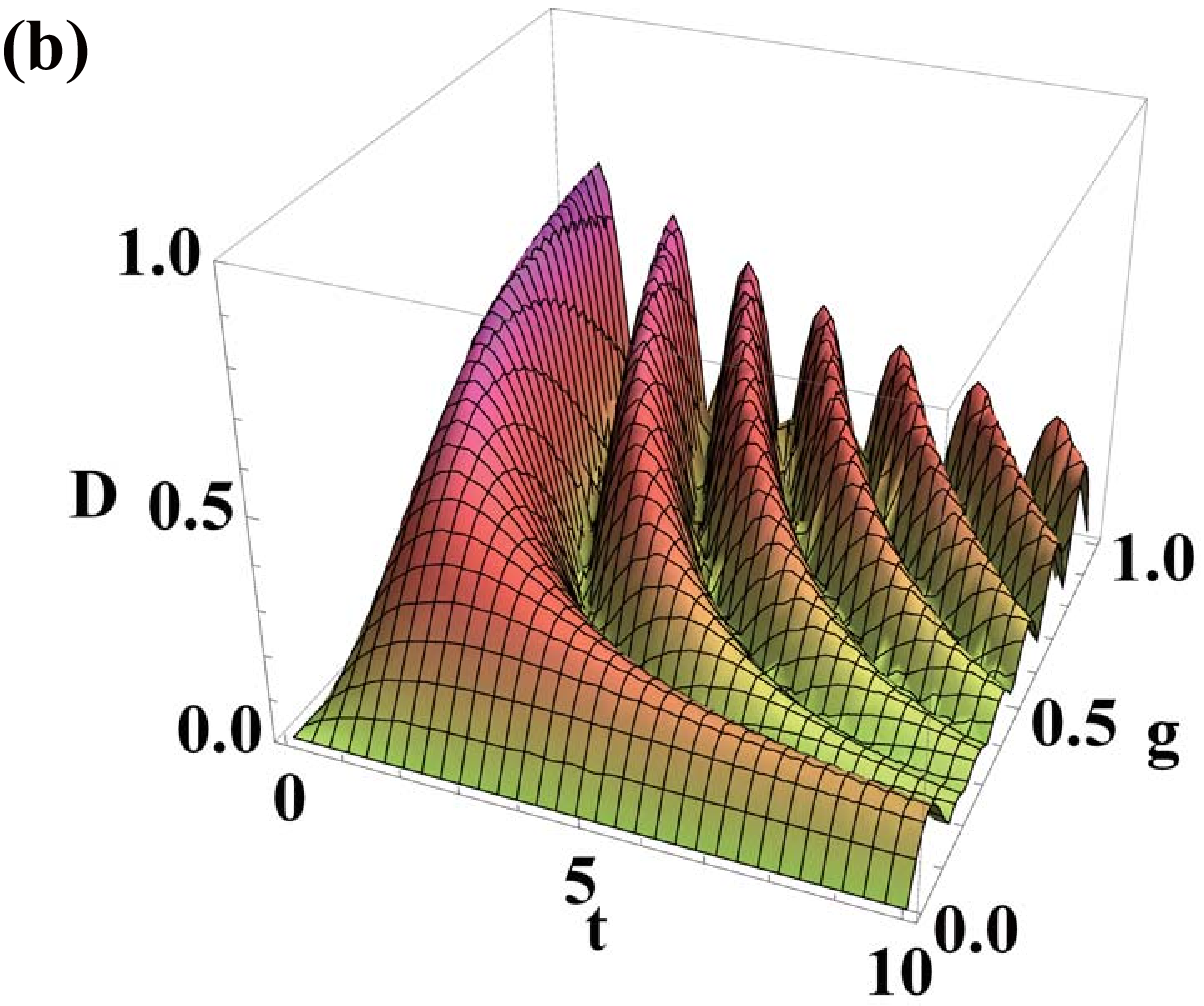}
\caption{(Color online) (a) Concurrence and (b) TDD as functions of $g$ and $t$ for initial EWL states (\ref{EWLPsi0}) with $r=2/3$ and $\theta=\phi=0$. Other parameters are set the same as Fig.~\ref{fig:fig4}.}
\label{fig:fig6}
\end{figure}
Next, the effect of $\theta$ of the initial EWL state on the nonclassical correlations including concurrence and TDD are also demonstrated in Fig.~\ref{fig:fig5}. It is found that non-Markovian revivals of nonclassical correlations can emerge by adjusting $\theta$.  Near $\theta=\pi/2$, both the concurrence and TDD exhibit Markovian decays as in Fig.~\ref{fig:fig4}. By contrast, in the region far away from $\theta=\pi/2$ where there is little amount of initial nonclassical correlations, both concurrence and TDD increase to their maximums at first and then decay with revivals. The difference between concurrence and TDD is that the concurrence possesses only single revival after ESD while the TDD can revival immediately several times without vanishing for a period of time. We would like to point out that the growth of nonclassical correlations at first and the non-Markovian revivals are induced by the dipole-dipole coupling $g$ between qubits $A$ and $B$. This can be concluded from Eqs.~(\ref{xsol1}) and (\ref{xsol2}) that when $g=0$ the qubits $A$ and $B$ are decoupled and under pure dephasing with constant diagonal elements $\rho_{ii}(t)=\rho_{ii}$(0) for $i=1,2,3,4$ and exponentially decaying counter-diagonal elements $|\rho_{14}(t)|=|\rho_{14}(0)|\exp(-2t/T_{S})$ and $|\rho_{23}(t)|=|\rho_{23}(0)|\exp(-2t/T_{S})$.
Therefore, it is easy to verify that the nonclassical correlations in the case of $g=0$ is under Markovian decay without any revival.
When qubits $A$ and $B$ are dipole-dipole coupled with $g>0$, then the diagonal elements $\rho_{22}(t)$ and $\rho_{33}(t)$ are no longer constant and the counter-diagonal elements $|\rho_{23}(t)|$ is not monotonic decreasing function versus $t$, in terms of which revivals of nonclassical correlations are possible. However, as has been shown in Fig.~\ref{fig:fig5}, non-Markovian revivals are still absent and the dephasing effect dominates the dynamics for highly correlated initial states. When the initial state is less correlated, non-Markovian revivals may be induced by the dipole-dipole coupling $g$. To see the effect of $g$ more explicitly, we consider the extreme case of initial state without any quantum correlation and plot concurrence and TDD as functions of $g$ and $t$ in Fig.~\ref{fig:fig6}. It is clearly seen that both concurrence and TDD can be induced followed with revivals and the number of revivals increases with $g$, which means the non-Markovian effect can be modulated by the dipole-dipole coupling $g$. Since the qubits $A$ and $B$ are under local dephasings, the nonclassical correlation must be induced by their dipole-dipole interaction. Besides, the difference between concurrence and TDD is that TDD can be induced immediately with both $g$ and $t$ while concurrence is suddenly created after certain finite time of evolution only for strong enough $g$.

\begin{figure}[!htb]
\centering
\includegraphics[angle=0,width=5.5cm]{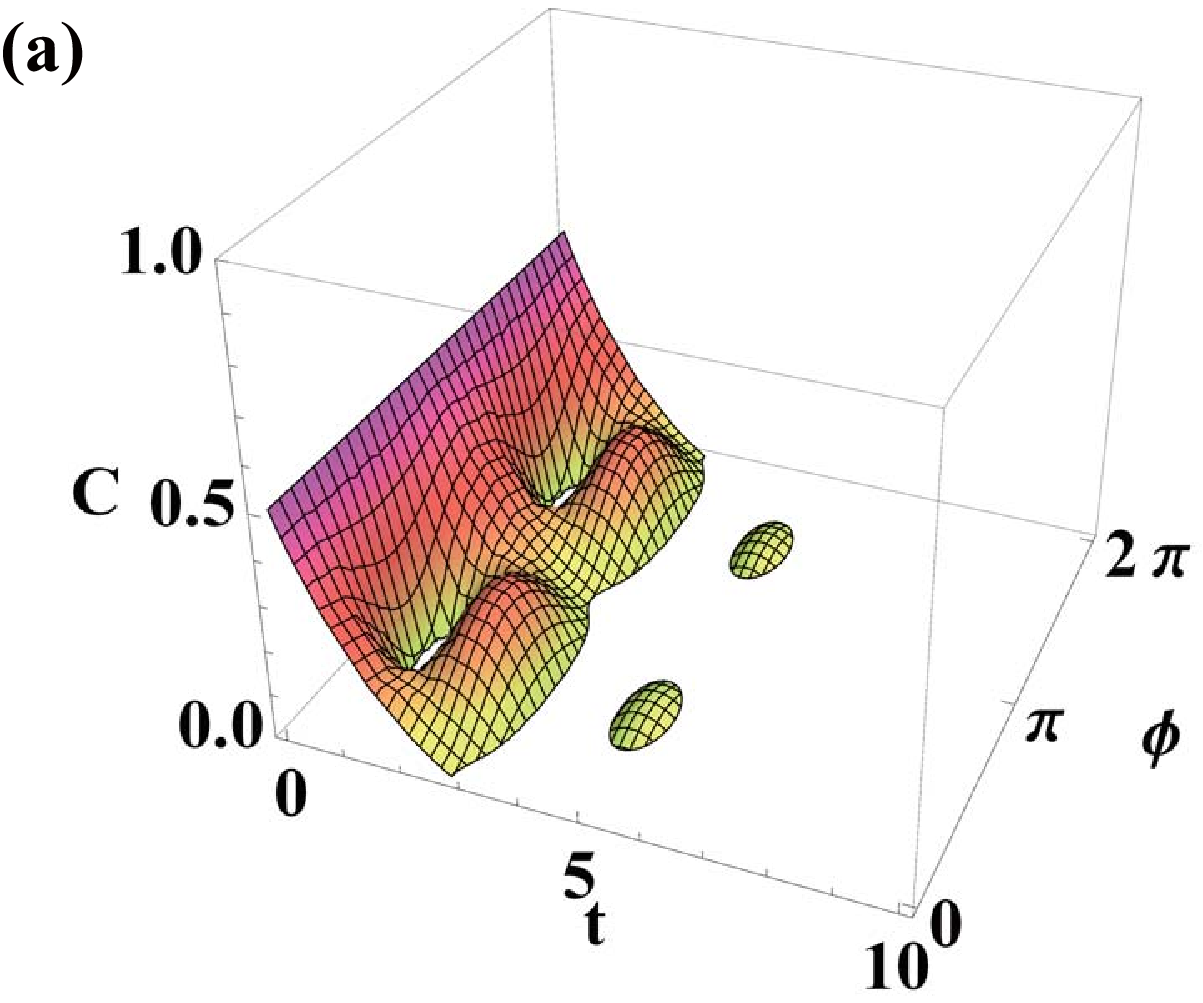}
\includegraphics[angle=0,width=5.5cm]{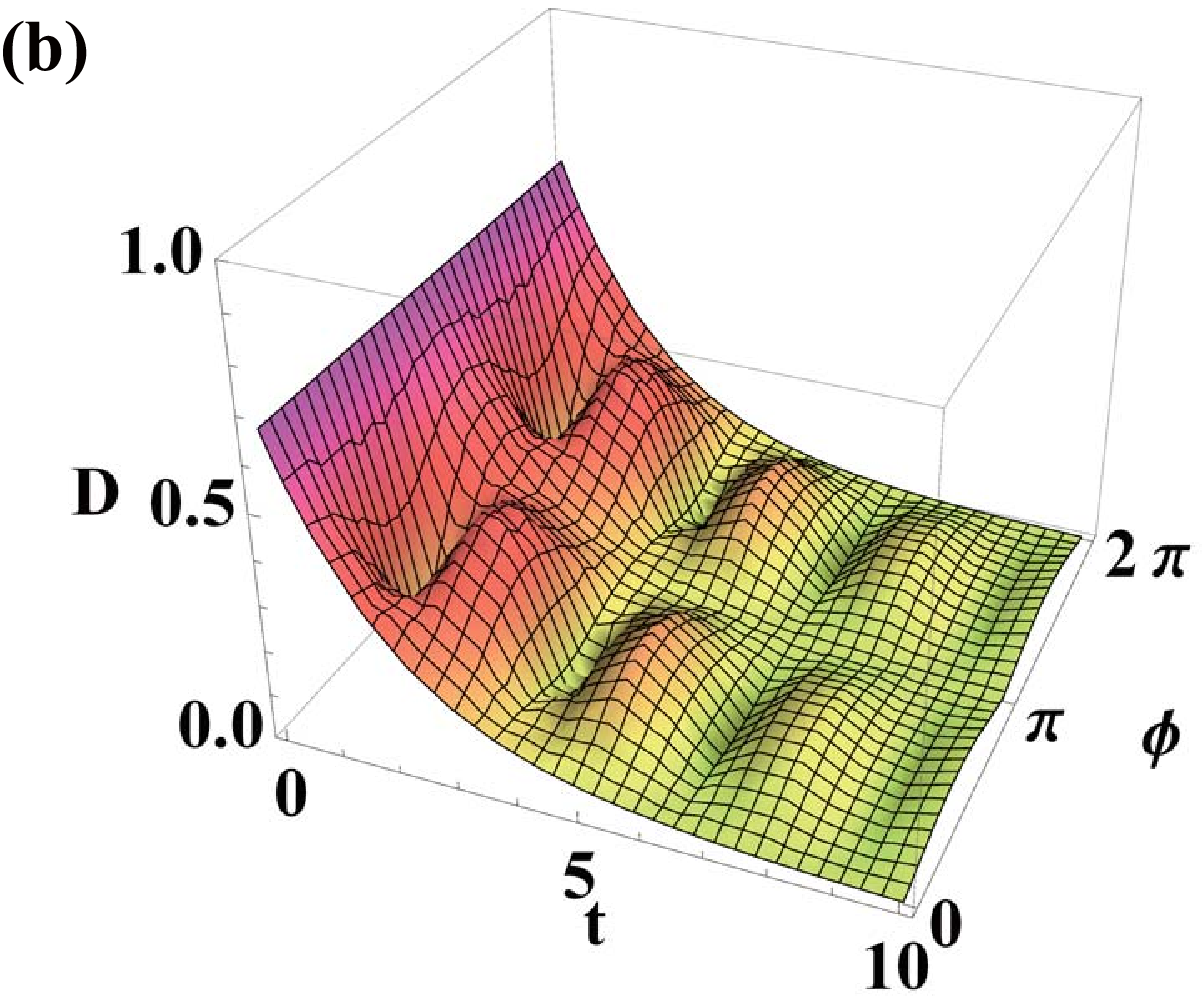}
\caption{(Color online) (a) Concurrence and (b) TDD as functions of $\phi$ and $t$ for initial EWL states (\ref{EWLPsi0}) with $r=2/3$ and $\theta=\pi/2$. Other parameters are set the same as Fig.~\ref{fig:fig4}.}
\label{fig:fig7}
\end{figure}
Finally, we discuss the effect of the relative phase $\phi$ in the initial $\Psi$-type EWL state $\rho^{(\Psi)}$ on concurrence and TDD.
It is seen From Eq.~(\ref{EWLPsi0}) that the initial amount of nonclassical correlations is independent on the phase $\phi$.
However, we can also see from Eq.~(\ref{xsol1}), where $\Re(\rho_{23}^{(\Psi)}(0))=\frac{r}{2}\sin\theta\cos\phi$ and $\Im(\rho_{23}^{(\Psi)}(0))=-\frac{r}{2}\sin\theta\sin\phi$ are contained in the elements $\rho_{22}^{(\Psi)}(t)$, $\rho_{33}^{(\Psi)}(t)$ and $|\rho_{23}^{(\Psi)}(t)|$ of $\rho^{(\Psi)}(t)$, that the dynamics of concurrence and TDD indeed depend on the relative phase $\phi$. To show this, we display the concurrence and TDD as functions of $\phi$ and $t$ in Fig.~\ref{fig:fig7}. It is clearly demonstrated that twice revivals of concurrence after ESD can be modulated when the relative phase is close to $\pi/2$ or $3\pi/2$. By contrast, several revivals of TDD without vanishing for a while always exist as long as the relative phase is not located at $0$ or $\pi$. Besides, local maximums can be obtained for both concurrence and TDD when the relative phase is exactly located at $\pi/2$ or $3\pi/2$. The physical mechanism of the dependence of nonclassical correlations on the relative phase $\phi$ is that quantum interference has been induced by the dipole-dipole coupling $g$ between qubits $A$ and $B$ in the dynamical evolution process, which can be easily checked that the influence of $\phi$ can be simply removed by setting $g=0$ in Eq.~(\ref{xsol1}).

\section{Conclusions}\label{sec:sec5}
In conclusion, we have investigated the dynamics of TDD for special $X$ states. By considering a dephasing two-qubit system initially prepared in a class of $X$ states with maximally mixed marginals, we find the necessary condition for the occurrence of freezing discord and compare it with that of entropic discord. It is shown that the condition of freezing TDD is much weaker than that of freezing entropic discord. Moreover, we employ a concrete model consisting of two interacting qubits coupled to independent reservoirs to illustrate these  dynamical properties of TDD. It is further found that the phenomena of freezing TDD exist even when the effect of dipole-dipole coupling between the two qubits is included, in which case the evolution state does not possess maximally mixed marginals. By increasing the dipole-dipole coupling $g$, the duration of freezing discord is prolonged in the sudden transition process while it is shortened in the double sudden changes process.

The differences between discord and entanglement are also analyzed by considering another class of $X$ states, i.e., the EWL states. The influences of parameters of the EWL state as well as the dipole-dipole coupling on the dynamics of nonclassical correlations are explored. We find that non-Markovian revivals and quantum interference can be induced by the dipole-dipole coupling.

\section*{Acknowledgments}
This work is supported by the Fundamental Research Funds for the Central Universities (Grant No.~JUSRP11405),
the natural science foundation of Jiangsu province of China (Grant Nos.~BK20140128 and BK20140167),
the National Natural Science Foundation of China (Grant Nos.~11274274 and~11247308)
and the National Natural Science Foundation of special theoretical physics (Grant No.~11347196).

\end{document}